\documentclass[a4paper,fleqn,usenatbib]{mnras}

\usepackage{newtxtext,newtxmath}

\usepackage[T1]{fontenc}

\usepackage{graphicx}	
\usepackage{amsmath}	
\usepackage{amssymb}	

\usepackage{enumerate}
\usepackage{epstopdf}
\usepackage{array}
\usepackage{xspace}
\usepackage{soul}

\newcommand{\qdep}{$Q_{\rm dep}$\xspace}
\newcommand{\lbol}{$L_{\rm bol}$\xspace}
\newcommand{\angstrom}{\textup{\AA}}
\newcommand{\nickel}{$^{56}\rm Ni$\xspace}

\title[SN Ia WLR I. Bolometric]{Type Ia supernovae have two physical width-luminosity relations\\
	and they favor sub-Chandrasekhar and direct collision models \\ I. Bolometric}

\author[Wygoda et. al.]{
	Nahliel Wygoda,$^{1,2}$\thanks{E-mail: nahliel.wygoda@weizmann.ac.il}
	Yonatan Elbaz$^{2}$
	Boaz Katz$^{1}$
	\\
	$^{1}$Dept. of Particle Phys. \& Astrophys., Weizmann Institute of Science, Rehovot 76100, Israel\\
	$^{2}$Dept. of Physics, NRCN, Beer-Sheva 84190, Israel\\
}

\pubyear{2018}

\date{Accepted XXX. Received YYY; in original form ZZZ}

\begin{document}
\label{firstpage}
\pagerange{\pageref{firstpage}--\pageref{lastpage}}
\maketitle

\begin{abstract}
While the width-luminosity relation (WLR) among type Ia supernovae (slower is brighter) has been extensively studied, its physical basis has not been convincingly identified. In particular, the 'width' has not been quantitatively linked yet to a physical time scale. We demonstrate that there are two robust fundamental time scales that 1. can be calculated based on integral quantities of the ejecta, with little dependence on radiation transfer modeling and 2. can be inferred from observations. The first is the gamma-ray escape time $t_0$, which determines the long-term evolution of the bolometric light curve and is studied in this Paper I. The second is the recombination time of $^{56}$Fe and $^{56}$Co, which sets the long-term color evolution of the emitted light and is studied in Paper II. Here we show that the gamma-ray escape time $t_0$ can be derived with $\sim 15\%$ accuracy from bolometric observations based on first principles. When applied to a sample of supernovae, the observed values of $t_0$ span a narrow range of $30-45$ days for the wide range of observed $^{56}\rm Ni$ masses $0.1M_{\odot}\lesssim M_{^{56}\rm Ni}\lesssim 1 M_{\odot}$. 
This narrow range of the gamma-ray escape time across the range of luminosities (a trivial WLR) is
consistent with central detonations and direct collisions of sub-Chandrasekhar mass white dwarfs (WDs) but not with delayed detonation models for explosions of Chandrasekhar mass WDs, which are therefore disfavored as the primary channel for the population of type Ia supernovae. Computer codes for extracting $t_0$ from observations and models and for calculating gamma-ray transfer in 1D-3D are provided.
\end{abstract}

\begin{keywords}
	radiative transfer -- Supernovae: Type Ia
\end{keywords}

\section{Introduction}

There is strong evidence suggesting that type Ia SNe, which comprise most of the observed SNe, are the result of thermonuclear explosions of WDs (see e.g. the recent reviews \citet{hn00,mmn14}). But what triggers the explosion in some (about $1\%$ of) WDs to produce type Ia SNe remains an open question.
Ideas that have been put forth can be separated based on the mass of the exploding WD, including 1. Chandrasekhar mass   models ($M_{ch}$ models): WDs close to the Chandrasekhar mass which are triggered by central heating due to continuing accretion (e.g. \citet{hf60,Arnett69,kho91}) 2. Sub-Chandrasekhar mass models (sub-$M_{ch}$ models): massive WDs with mass of about  $\approx 1M_{\odot}$ which are triggered by the explosion of a helium shell during accretion (e.g. \citet{wtw86,Livne90,sb14,s+17}) or by a violent merger with another WD (e.g. \citet{Pakmor10}). 3. Collisions: direct collisions of typical WDs with masses $\approx 0.5-1 M_{\odot}$ (e.g. \citet{rkgr09,kd12,kk13,dkkp15}).

Most of the observed data on type Ia supernovae consists of spectra and light curves in the optical regime. It has long been realized that type Ia supernovae span a significant range in luminosities (peak luminosity varying in the range $10^{42}-10^{43}\rm ergs/s$) which are correlated with the timescales for the rise and decline of the light in the different bands (e.g. \citet{p77,phil93,phil05}). Quite generally, brighter type-Ia tend to evolve more slowly. This so called width-luminosity relation (WLR) is crucial for using type Ia's as standard candles.


The brightness of a type Ia is mostly set by the amount of $^{56}$Ni produced. The observed range of brightness implies that the progenitors have a spread in $^{56}\rm Ni$ yield in the range $0.1\rm M_{\odot}\lesssim M_{^{56}Ni}\lesssim 1 M_{\odot}$. But which feature of the explosion sets the evolution time-scale and what does the correlation teach us about the ejecta? It has become clear that both the temporal evolution of the bolometric light curve (e.g. \citet{pi00a,pi01}) and the color (e.g. \citet{kw07}) are important and may depend on the properties of the ejecta. A particularly important question for making progress with identifying the explosion mechanism is the relation to the total ejected mass. Is the sequence hinting to a spread of masses (e.g. \citet{phil93, pi00a,kk13, s+14,b+17})? or is it a result of composition variations in explosions that all eject $M_{ch}$ (e.g. \citet{h+96,pi01,m01,kw07,h+17})?  The results presented here provide evidence that the range of type Ia's is not likely to originate from  $M_{ch}$ explosions. 

It is important to note that while some studies investigate the entire range of observed brightness (e.g. \citet{s+10, b+17,h+17}), other studies ignore the faint end and focus only on the bright part of the relation (e.g. \citet{pi01,m01,kw07}), assuming that there is a separate mechanism causing the low end. Yet other studies \citep{dhawan17} have shown that the observation of separate subclasses at the low luminosity end might point to the presence of several explosion mechanisms. Given the continuous distribution of type Ia photometric (e.g. \citet{phil05}) and spectroscopic properties (e.g. \citet{n+95,b+09}), and the success of some calculations in producing the full range of brightness (e.g. \citet{h+96,s+10,kk13}) it is possible that the majority of type Ia's across the brightness range originate from a single underlying mechanism with continuous parameters. In this paper we study the WLR across the entire brightness range.

Studies of the emission from type Ia face the following significant challenges: First, we don't know the explosion scenario, allowing a very large parameter space of possibilities to study. Second, given a specific scenario, the calculation of the explosion and radiation transfer requires the use of approximations which are not clearly valid to sufficient accuracy. A nice demonstration of the challenge in radiation transfer modeling is the recent 1-D calculations of two of the most popular models by three different groups this year \citep{s+17,b+17,h+17},  which obtain contradictory results of the WLR for essentially the same explosion scenarios: compare the core detonation sub-Chandra models in  figure 13 of \citet{s+17} to those in figure 5 of \citet{b+17} and the delayed detonation Chandra models in figure 5 of \citet{b+17} to those in figure 8 of \citet{h+17}. 

In order to make progress, we believe it is essential to identify \textbf{robust and quantitative} features of models and observations which are insensitive to the uncertainties of radiation transfer.

Energy conservation can be used to bypass detailed radiation transfer if the bolometric light curve can be accurately inferred from observations.  One robust parameter which can be inferred from the bolometric light curve is the mass of $^{56}\rm Ni$, which largely sets the brightness scale of the light curve. Another parameter which would be very useful to infer is the total mass of the ejecta $M_{\rm tot}$ \citep[e.g.][]{s+14}. While there is no known way to infer the total mass in a model-independent way, the total column density (weighted by $^{56}$Ni) can be inferred from the shape of the light curve at late times \citep[e.g.][]{j99}. The total column density sets the average optical depth for gamma-rays at late times and is parametrized by the gamma-ray escape time $t_0$ (defined such that at late times, $t\gg t_0$, the fraction of energy in gamma-rays that is deposited in the ejecta is $t_0^2/t^2$, see \S\ref{sec:physics}). $t_0$ is a fundamental time scale in the evolution of the light curve and in particular sets the slope of the decline of the bolometric light.

In this Paper I we present a new method  to derive the total $^{56}\rm Ni$ mass and the gamma-ray escape time $t_0$ which is simple and directly based on energy conservation. We show that these parameters are robust, and quantify the estimated uncertainty in their inference. We apply this method to bolometric light curves from the literature to study the physical relationship between $t_0$ and $^{56}$Ni, which we call the "bolometric WLR" (and is different from the "usual" WLR which refers to various bands of the spectrum). In \cite{wygoda18} (hereafter Paper II) we study a second physical time-scale which is fundamental to the WLR, namely the time of recombination of iron group elements from doubly ionized to singly ionized. We show that it can be robustly calculated theoretically and inferred from observations using the distinctive brake in the color evolution of type Ia's around 30 days after peak \citep[e.g.]{burns14,p77}. Any successful model needs to agree with the two WLRs obtained from the bolometric and the color time scales. By comparing the results of representative models of Chandrasekhar, sub-Chandrasekhar and direct collision scenarios, we find that all agree with the color WLR but only the sub-Chandrasekhar and direct collisions are consistent with the bolometric WLR. Chandrasekhar-mass models are consistent with bright type Ias and are inconsistent with faint type Ia's.

This Paper I is organized as follows. In \S\ref{sec:physics}, a method to infer $^{56}\rm Ni$ mass and $t_0$ from observed light-curves is derived based on energy conservation arguments. In \S\ref{sec:simulations}, detailed 1-D radiation transfer calculations are used to validate the method for representative models and study the sensitivity to the quality of the observed light-curve and level of approximation. In \S\ref{sec:observations}, the method is applied to extract the parameters from samples of tens of observed SNIa. In \S\ref{sec:models}, the observed ejecta parameters are compared to Chandrasekhar, sub-Chandrasekhar and direct collision scenarios. The results are summarized in \S\ref{sec:discussion}. The analysis files in python (Jupyter notebooks) and matlab, as well as a c-based gamma-ray Monte Carlo code are provided with the manuscript and are described in appendix \S\ref{sec:files}.

\section{The gamma-ray escape time and $^{56}$Ni mass can be directly extracted from bolometric light curves}\label{sec:physics}

In this section we describe how the $^{56}$Ni mass and gamma-ray escape time $t_0$ can be extracted from bolometric UVOIR light curves. 

\subsection{Energy conservation connects the deposition rate to the bolometric light curve}
The radioactive decay chain of $^{56}\rm Ni$ emits energy in the form of $\gamma$-rays and positrons which traverse the ejecta. At any given time $t$, the gamma-rays and positrons deposit some of their energy in the ejecta plasma at a total rate of \qdep which is essentially instantaneously converted to UVOIR radiation. The total energy deposition rate $Q_{\rm dep}(t)$ can be accurately calculated using a gamma-ray transport code or approximated analytically (see \S\ref{sec:simplified_qdep}). It is important to emphasize that the physics of gamma-ray transfer is well understood. The dominant interaction of gamma-rays with the plasma is Compton collisions with all electrons and is practically independent of temperature and ionization. Each gamma-ray photon can be calculated separately using Monte-Carlo simulations. The total deposition rate for any 3D model ejecta can be calculated easily and accurately (a Monte Carlo code is attached to the manuscript and described in appendix \S\ref{sec:files}). The question we address here is how to relate the total deposition rate \qdep to the UVOIR bolometric light curve \lbol. 

At sufficiently late times after explosion, the diffusion time becomes much shorter than the dynamical time and energy deposited is immediately emitted:
\begin{equation}\label{eq:lateQL}
L_{\rm bol}(t)=Q_{\rm dep}(t)~~\rm{for}~~t\gg t_{\rm peak}.
\end{equation}

At early times the UVOIR radiation is trapped and the deposition cannot be directly related to the emission. As the radiation diffuses through the ejecta its energy is diluted adiabatically due to the expansion and the integrated emission is smaller than the integrated deposition
$\int dt L_{\rm bol}(t)<\int dt Q_{\rm dep}(t)$. Since the expansion is adiabatic to an excellent approximation, the time weighted energy in the radiation is conserved and we have $tE_{\rm rad}(t)+\int dt t L_{\rm bol}(t)=\int dt t Q_{\rm dep}(t)$, where $E_{\rm rad}(t)$ is the total energy in UVOIR radiation in the ejecta at time $t$ \citep{kkd13}. At late times there is little trapped radiation and to a good approximation we have
\begin{equation}\label{eq:qtdt}
\int_0^{t} dt' t' L_{\rm bol}(t')=\int_0^t dt' t' Q_{\rm dep}(t')~~\rm{for}~~t\gg t_{\rm peak}.
\end{equation}

Thus, given an ejecta model and the corresponding \qdep, the model can be compared to actual observations using Eqs. \eqref{eq:lateQL} and \eqref{eq:qtdt}. In particular, it is useful to compare the ratios of the two equations \citep[][see also figure \ref{fig:bolo_fit_t0}]{kk13}:
\begin{equation}\label{eq:ratio}
\frac{L_{\rm bol}(t)}{\int_0^{t} dt' t' L_{\rm bol}(t')}=\frac{Q_{\rm dep}(t)}{\int_0^t dt' t' Q_{\rm dep}(t')}~~\rm{for}~~t\gg t_{\rm peak}.
\end{equation}
The ratios in equation \eqref{eq:ratio} do not depend on the distance or the total amount of $^{56}$Ni (both numerator and denominator scale linearly with the $^{56}$Ni fraction) and thus provide direct information on the gamma-ray escape fraction \citep{kk13}.

\subsection{The deposition rate is set by the $^{56}$Ni mass and the gamma-ray escape time $t_0$} \label{sec:simplified_qdep}
The rate of energy deposition in the ejecta from $\gamma$-rays can be precisely expressed analytically in two limits: 1. At early times, when the ejecta is dense enough that all $\gamma$-rays are trapped and their energy is fully deposited in the ejecta, the gamma-ray deposition fraction $f_{\rm{dep},\gamma}$ is:
\begin{equation}\label{eq:dep_early}
f_{\rm{dep},\gamma}=1~~\rm{at}~~ t\lesssim t_{\rm peak}.
\end{equation}
2. At late times the ejecta becomes optically thin to gamma-rays and each gamma-ray has a small chance to experience a Compton collision (and negligible chance to have more than one collision). The deposition is thus accurately accounted for by an effective opacity $\kappa_{\rm eff}$ which is calculated by averaging the Klein-Nishina corrected Compton cross section $\Sigma_C(E_n)$ and average fractional energy loss per scattering $<dE>/E$ over each of the (discrete) emitted gamma-ray energies $E_n$ \citep[including the positron annihilation line][]{s+95,j99}
\begin{equation}
\kappa_{\rm eff}=Y_e\sum_{\gamma~\rm lines}f_n\frac{\sigma_C(E_n)}{m_p}\frac{\langle dE\rangle_{n}}{E_n}\approx 0.025\,\rm cm^2/gr
\end{equation}
where $f_n$ is the fraction of energy emitted in the line $n$, and it was assumed that there are two baryons per electron ($Y_e=0.5$).  The gamma-ray deposition fraction at late times is given by 
\begin{equation}\label{eq:dep_late}
f_{\rm{dep},\gamma}=\kappa_{\rm eff}\langle\Sigma\rangle_{\rm Ni}=\frac{t_0^2}{t^2}~~\rm{at}~~ t\gg t_{\rm peak},
\end{equation}
where $\langle\Sigma\rangle_{\rm Ni}\propto t^{-2}$ is the average column density of the ejecta as seen by the $^{56}$Ni elements:
\begin{equation}\label{eq:SigmaNi}
\langle\Sigma\rangle_{\rm Ni}=\int \frac{d^3x}{M_{\rm Ni56}}~\rho_{\rm Ni56}({\bf x})\int \frac{d{\bf\hat\Omega}}{4\pi} \int_0^{\infty} ds~\rho({\bf x}+s{\bf \hat \Omega})
\end{equation}
where $\rho_{\rm Ni56}$ is the density of $^{56}$Ni and $\rho$ is the total density. The escape time $t_0$ is given by
\begin{equation}\label{eq:t0exp}
t_0=\sqrt{\kappa_{\rm eff}t^2\langle\Sigma\rangle_{\rm Ni56}}
\end{equation}
and is independent of time (assuming homologous expansion). At times $t\gtrsim t_0$ a significant fraction of the emitted gamma-rays escape the ejecta without depositing their energy. $t_0$ can be easily calculated for any ejecta using equation \eqref{eq:t0exp}. A Python based code to calculate $t_0$ for 1-D ejecta and c-based codes for calculating the gamma-ray transport and $t_0$ for 1-D,2-D and 3-D are provided in the attached material as described in appendix \S\ref{sec:files}.

A useful interpolation for the deposition function between the two exact limiting values  in Eqs. \eqref{eq:dep_early} and \eqref{eq:dep_late} is \citep[e.g.][]{s+14}
\begin{equation}\label{eq:fdepinterp}
f_{\rm{dep},\gamma}(t)\approx 1-e^{-(\frac{t_0}{t})^2}.
\end{equation}
The total deposition rate at any time can be thus approximated by:
\begin{equation}
Q_{\rm dep}(t)\approx Q_{\gamma}(t)\cdot(1-e^{-(\frac{t_0}{t})^2})+Q_{\rm pos}(t)
\label{eq:qdep}
\end{equation}

where $Q_{\gamma}$ and $Q_{\rm pos}$ are the total energy release rate of gamma-rays and positron kinetic energy respectively and are given by:
\begin{align}
Q_{\gamma}(t)&=N_{\rm Ni}\cdot \left[\frac{Q_{\rm Ni}}{\tau_{\rm Ni}}\cdot e^{-\frac{t}{\tau_{\rm Ni}}}+\frac{Q_{\rm Co, \gamma}}{\tau_{\rm Co}-\tau_{\rm Ni}}\cdot \left(e^{\frac{-t}{\tau_{\rm Co}}}-e^{\frac{-t}{\tau_{\rm Ni}}}\right)\right]\cr
&=\frac{M_{\rm Ni56}}{M_{\odot}}\left[6.45~e^{-\frac{t}{8.76\rm d}}+1.38~e^{-\frac{t}{111.4\rm d}}\right]\times 10^{43}\rm ergs/s
\label{eq:qgamma}
\end{align}
and
\begin{align}
Q_{\rm pos}(t)&=N_{\rm Ni}\cdot\frac{Q_{\rm Co, pos}}{\tau_{\rm Co}-\tau_{\rm Ni}}\cdot \left(e^{\frac{-t}{\tau_{\rm Co}}}-e^{\frac{-t}{\tau_{\rm Ni}}}\right)\cr
&=4.64\frac{M_{\rm Ni56}}{M_{\odot}}\left[-e^{-\frac{t}{8.76\rm d}}+e^{-\frac{t}{111.4\rm d}}\right] \times10^{41}\rm ergs/s\cr
\label{eq:qpos}
\end{align}
where $N_{\rm Ni}$ is the number of synthesized $^{56}$Ni nuclei, the lifetimes of $^{56}\rm Ni$ and $^{56}\rm Co$ are $\tau_{\rm Ni}=8.76$ days and $\tau_{\rm Co}=111.4$ days, respectively \citep{junde99}, the mean $\gamma$-ray energies per decay are $Q_{\rm Ni}=1.73$ Mev, $Q_{\rm Co, \gamma}=3.57$ Mev \citep{as88} and  $Q_{\rm Co, pos}=0.12$ Mev.
Note that at very late times of hundreds of days past explosion positrons may escape the ejecta [see e.g. constraints by \citet{axelrod80} and \citet{ktsr14}].

\subsection{Extracting $t_0$ and $^{56}Ni$ from light curves}\label{sec:t0MNifitting}
\begin{figure*}
	\includegraphics[width=\columnwidth]{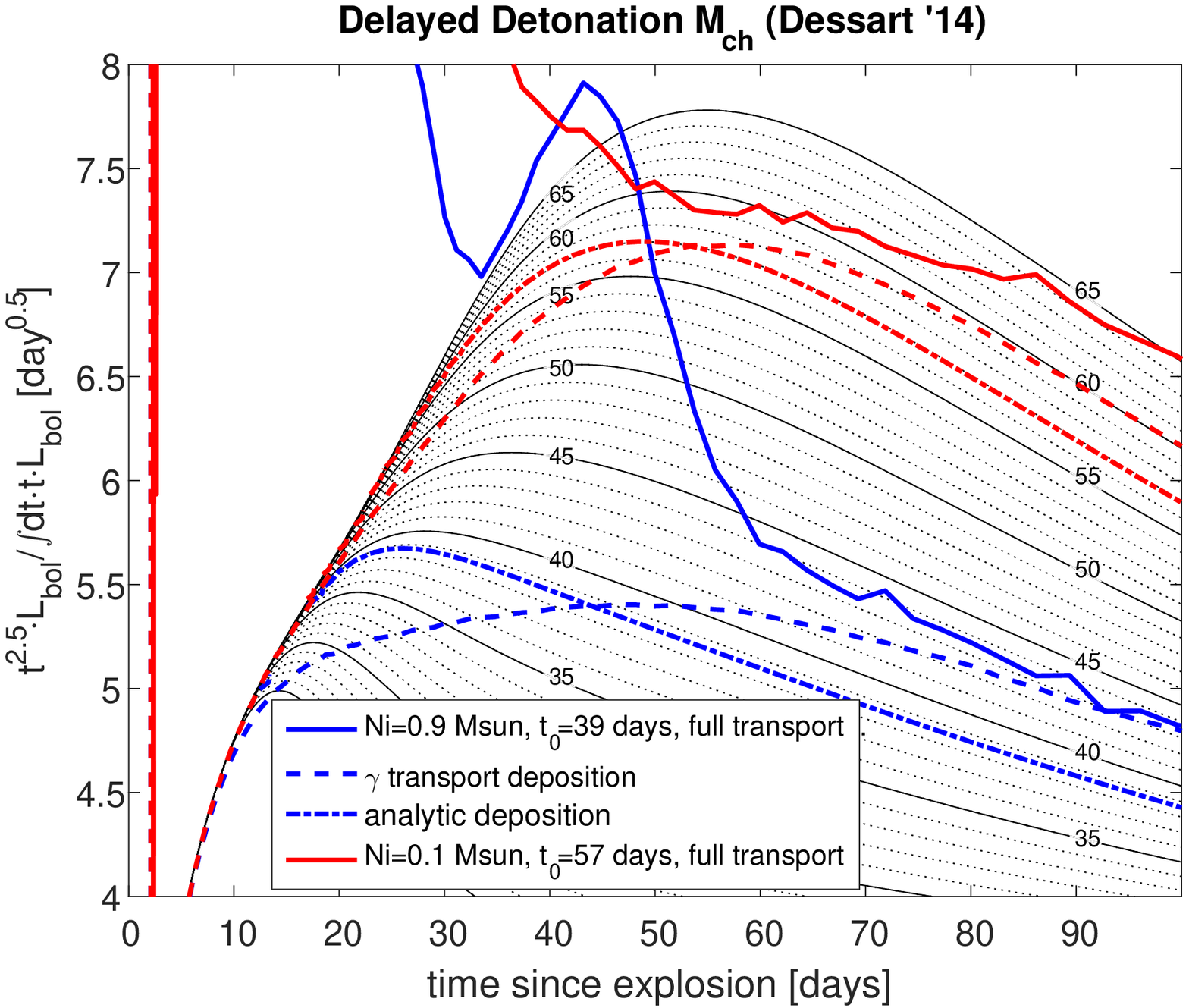}
	\hspace{2mm}
	\includegraphics[width=\columnwidth]{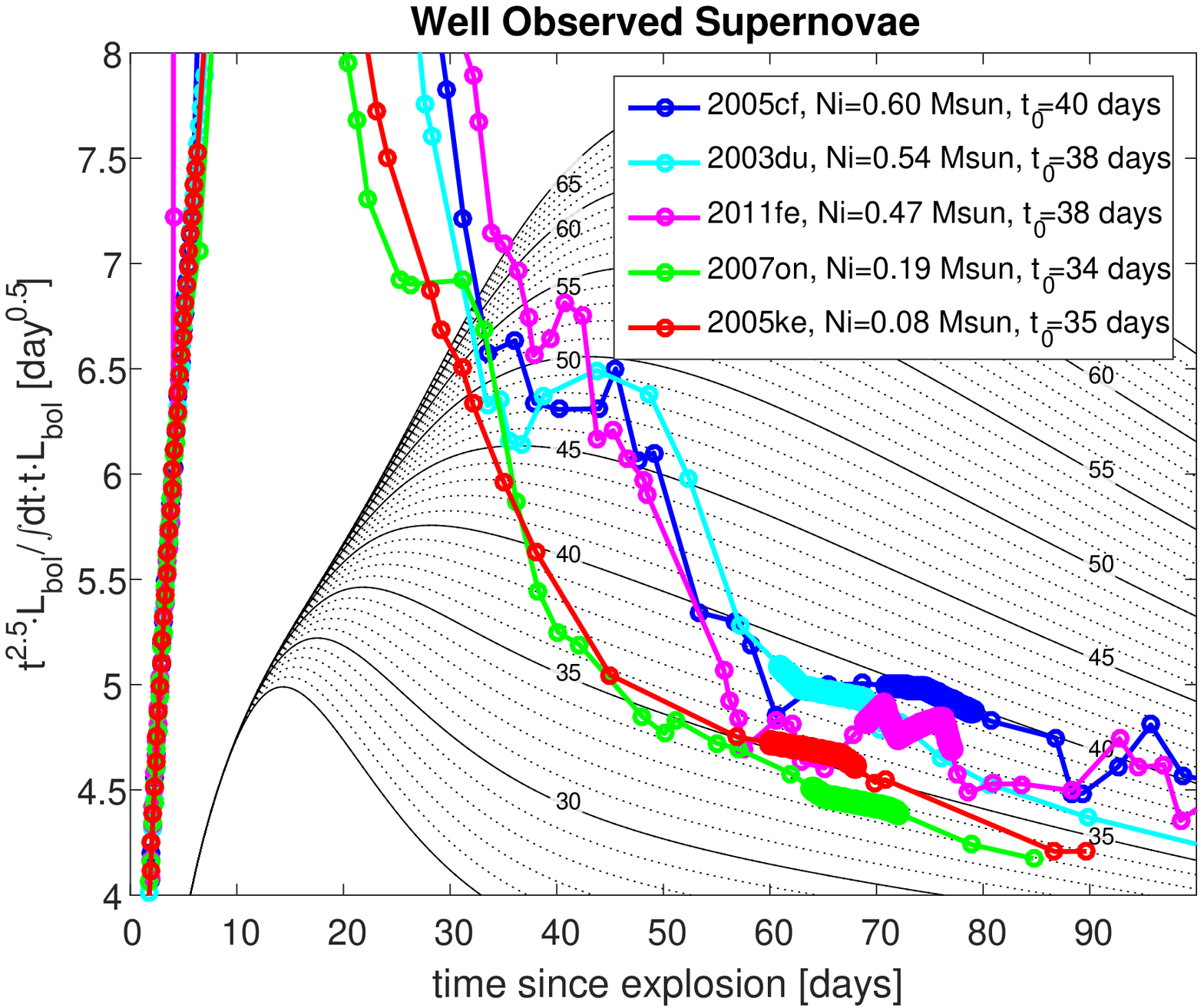}
	\vspace{2mm}	
	\includegraphics[width=\columnwidth]{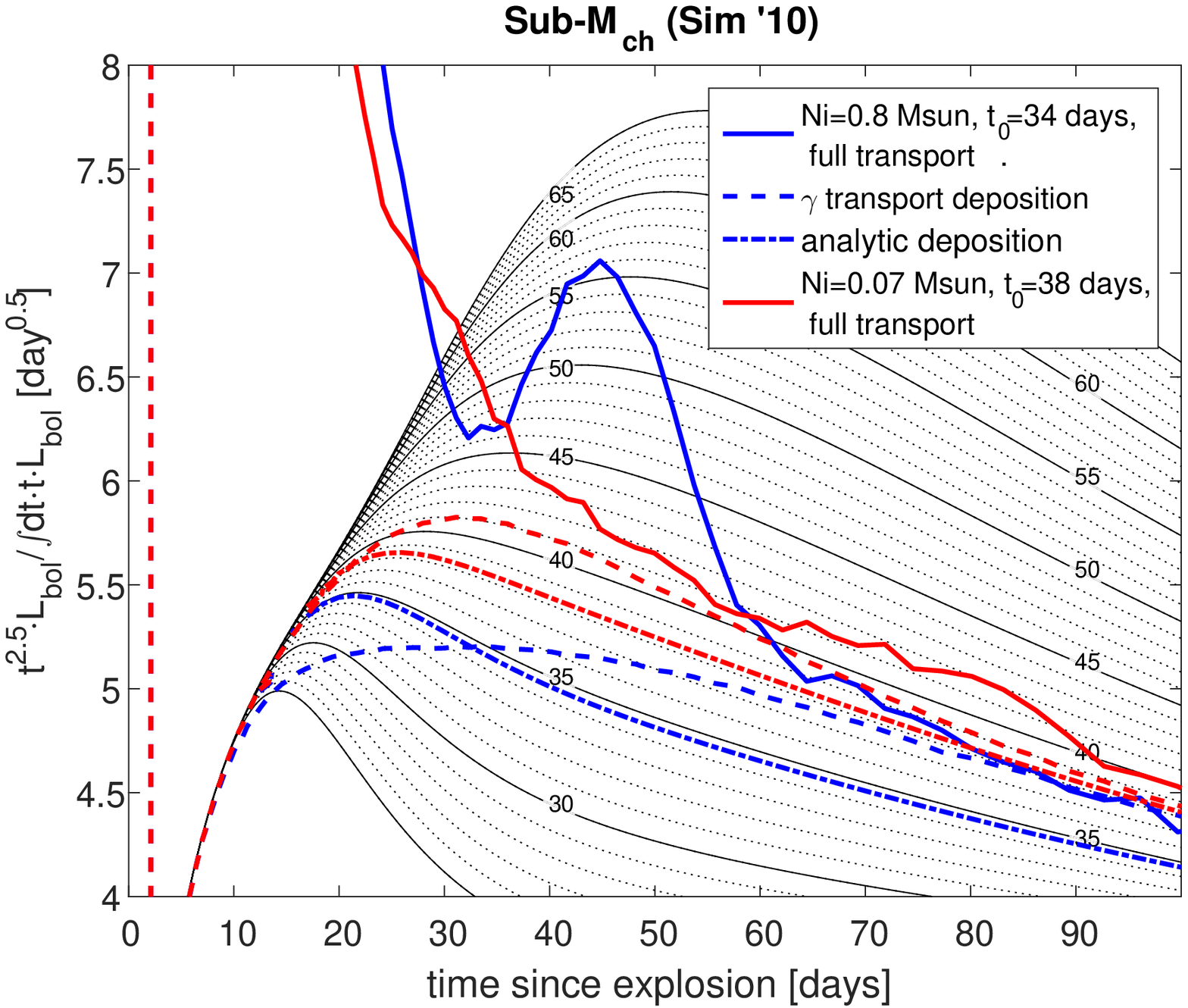}
	\hspace{2mm}
	\includegraphics[width=\columnwidth]{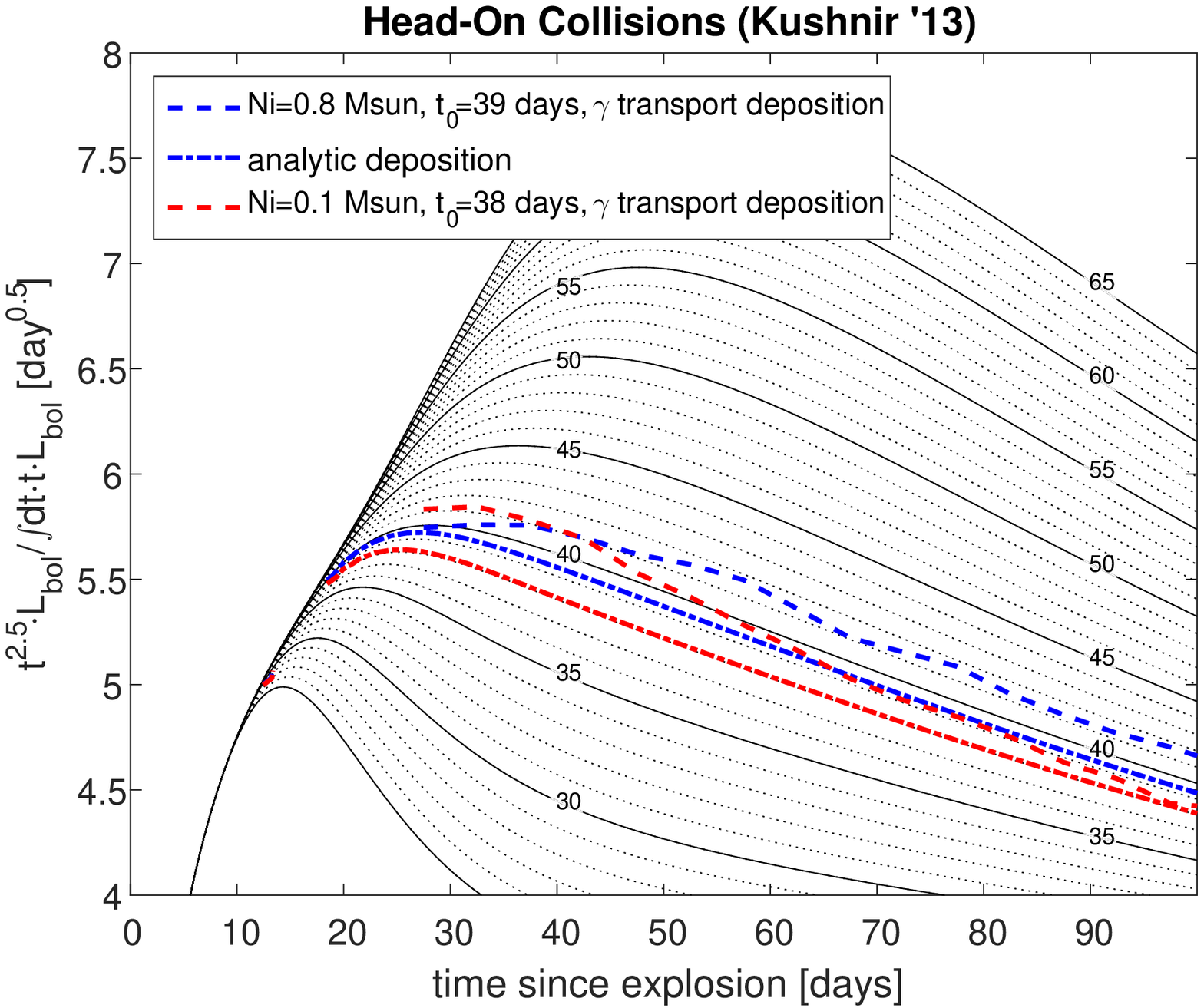}
	\caption{\textbf{Extracting the gamma-ray escape time $t_0$ from bolometric light curves.} Top right panel: The ratios $L_{\rm bol}/\int t dt L_{\rm bol}$ (multiplied by $t^{2.5}$ for convenience) are shown in solid lines for a sample of well observed supernovae: 2005cf \protect\citep[blue,][]{w+09}, 2003du \protect\citep[cyan,][]{s+07},  2011fe \protect\citep[magenta,][]{m+15}, 2007on \protect\citep[green,][]{phil12} and 2005ke \protect\citep[red,][]{phil12}. Analytic ratios of the deposition $Q_{\rm dep}/\int t dt Q_{\rm dep}$ are shown for a range of $t_0$ values using equations (\ref{eq:ratio}, \ref{eq:qdep}-\ref{eq:qpos}) in black and dotted lines with the values of $t_0$ labeled on the lines. As can be seen by comparing the observed ratios to the analytic curves at late times $t\gtrsim 60$ days all these supernovae have $t_0$ in the range of $35-40$ days. The thicker region on the lines marks the time range in which the value of $t_0$ used for our analysis was inferred. Top left and two bottom panels: the luminosity ratio functions obtained for synthetic light curves calculated for representative models  - $M_{ch}$ delayed detonation models DDC0 (blue) and DDC25 (red) from \protect\cite{dbhk14}, sub-$M_{ch}$ central detonations of WDs with masses $1.15\rm M_{\odot}$ (blue) and $0.88\rm M_{\odot}$ \protect\citep[red][]{s+10} and 2D models of equal mass head on collisions of $0.9\rm M_{\odot}$ WDs (blue) and $0.5\rm M_{\odot}$ WDs (red) from \protect\cite{kk13}.  Models with high (blue) and low (red) yields of $^{56}\rm Ni$ are presented (corresponding to the colors of the bright 2005cf and the faint 2005ke in the top right panel). The ratios $L_{\rm bol}/\int t dt L_{\rm bol}$ for the full radiation transfer calculations are shown in solid lines (except for the 2D collision models, for which the radiative transfer simulations have not been preformed yet). The ratios $Q_{\rm dep}/\int t dt Q_{\rm dep}$ are shown based on $\gamma$-ray transport calculations (dashed) and analytically (dash-dotted) using equations (\ref{eq:ratio}, \ref{eq:qdep}-\ref{eq:qpos}) for all models. 	
		The matlab file that created the top right panel, a python file for extracting $t_0$ and  $M_{^{56}Ni}$ from bolometric light curves, and python and c files for calculating $t_0$ and $Q_{\rm dep}$ for models are attached to the manuscript as described in appendix \S\ref{sec:files}.
	}
	\label{fig:bolo_fit_t0}
\end{figure*}

The luminosity ratio $L_{\rm bol}(t)/\int_{0}^{t} L_{\rm bol}(t')t'dt'$ can be used to extract $t_0$ using  Eqs. \eqref{eq:ratio} and (\ref{eq:qdep}-\ref{eq:qpos}). In the top right panel of figure \ref{fig:bolo_fit_t0} the luminosity ratios of five well observed supernovae from the literature that include UV, optical and IR  observations are shown. For convenience the luminosity ratios are multiplied by $t^{2.5}$, where $t$ is the estimated time since explosion in days. The expected ratios for a range of $t_0$ values, as given by Eqs. \eqref{eq:ratio} and (\ref{eq:qdep}-\ref{eq:qpos}), are shown in black lines with labels indicating the corresponding values of $t_0$ . As can be seen in the figure, the light-curves of all five supernovae converge beyond $\sim 60$ days to the analytic expectations that correspond to $t_0$ values around $35-40$ days \footnote{We define $t_0$ as the average value within the 10-day bin that has the smallest scatter of $t_0$ values in the range 55-80 days after explosion. The assigned value of the nickel mass is interpolated as the average value over the same time range given this defined $t_0$. Varying these arbitrary definitions over a reasonable range leads to negligible modifications in the assigned $t_0$ and nickel mass values.}.
Note that the presented supernovae include both bright and faint events. The uniformity of $t_0$ values across the type Ia luminosity range is confirmed for bigger samples (with less quality) in \S\ref{sec:observations}. Once $t_0$ is determined, the $^{56}$Ni can be inferred by comparing the amplitude of the light curve to the deposition function calculated with $t_0$ using Eqs. (\ref{eq:qdep}-\ref{eq:qpos}). The values of $^{56}$Ni extracted in this way are indicated in the legend.
The matlab file that created the top right panel and a python file for extracting $t_0$ and  $M_{^{56}Ni}$ from bolometric light curves are attached to the manuscript as described in appendix \S\ref{sec:files}.

\section{Radiation transfer simulations of representative models}\label{sec:simulations}
Equations \eqref{eq:ratio} and (\ref{eq:qdep}-\ref{eq:qpos}) are next compared to results of numerical simulations of both the $\gamma$-rays and the UVOIR transport through the ejecta of representative models of Chandrasekhar, sub-Chandrasekhar and the collision scenarios (for the latter, only $\gamma$-rays transport was performed, while the full, angle dependent light curves, will be studied in a forthcoming publication).

\subsection{Radiation transfer code}
Radiation transfer is calculated using URILIGHT, a Monte-Carlo code based on the approximations used in the SEDONA program (\cite{ktn06} and references therein). This simulation is based on several physical approximations: 1) homologous expansion. 2) expansion opacities with optical depths using the Sobolev approximation. 3) Local thermodynamic equilibrium (LTE) is assumed for calculating the ionization and excitation states. The atomic line data for the bound-bound transitions, which constitutes the main and most important opacities for these problems, are taken from \cite{kurucz95}. A detailed description of this program and comparisons to previously published radiative transfer codes for several benchmark problems are presented in Paper II. 

\subsection{Representative ejecta models}\label{sec:repmodels}
We present here the deposition function and bolometric light curves of 4 1D ejecta representing extremes of bright and dim SNIa in Chandrasekhar and sub-Chandrasekhar explosion models. The deposition functions for two 2D ejecta that result from extremes of 2D simulations of head-on collisions are also presented.  For the Chandrasekhar models, the delayed detonation models DDC0 and DDC25 from \citet[profiles provided by the authors]{dbhk14}, corresponding respectively to $^{56}\rm Ni$ masses of 0.86 and 0.12 $M_{\odot}$ were used. For the sub-Chandrasekhar models, the ejecta profiles generated from simple central explosions of sub Chandrasekhar mass WDs (1.15 and 0.88 $M_{\odot}$) from \citet[profiles provided by the authors]{s+10}, which have $^{56}\rm Ni$ masses of 0.81 and 0.07 $M_{\odot}$, were used. For the collision models, head on collisions of equal mass $0.9\rm M_{\odot}$ WDs and of equal mass $0.5 \rm M_{\odot}$ WDs from \cite{kk13}, which have $^{56}\rm Ni$ masses of 0.79 and 0.11 $M_{\odot}$, were used.

In the top left and bottom left panels of figure~\ref{fig:bolo_fit_t0} the calculated synthetic luminosity ratios (solid blue and red) of the 1D Chandrasekhar and sub-Chandrasekhar mass explosions, for which the full radiation transfer was calculated, are shown. Full radiation transfer was not calculated for the 2D collision models in the bottom right panel. The deposition ratios calculated using Monte Carlo gamma-ray transfer are shown for all models (dashed blue and red). As can be seen, beyond about  $60$ days after explosion (earlier for the faint models) the ratios of the bolometric luminosities converges to that of the deposition, given by Eq. \eqref{eq:ratio}, to an accuracy of a few percents (a bump with higher error appears at ~130 days in the high luminosity models but we do not consider times $>100$ days here where the LTE approximation probably breaks down). Note the bump in the luminosity ratios around $50$ days for the brighter events which is due to the IR second maximum and is absent in the lower luminosity models. The analytic deposition ratios calculated using Eqs. (\ref{eq:qdep}-\ref{eq:qpos}) are shown for each model in dashed dotted blue and red lines and for a range of $t_0$ values in black lines. As can be seen, beyond about 60 days after explosion, the analytic estimates agree with the deposition and luminosity ratios to an accuracy better than $10\%$.

\subsection{error budget}
\begin{figure}
	\includegraphics[width=\columnwidth]{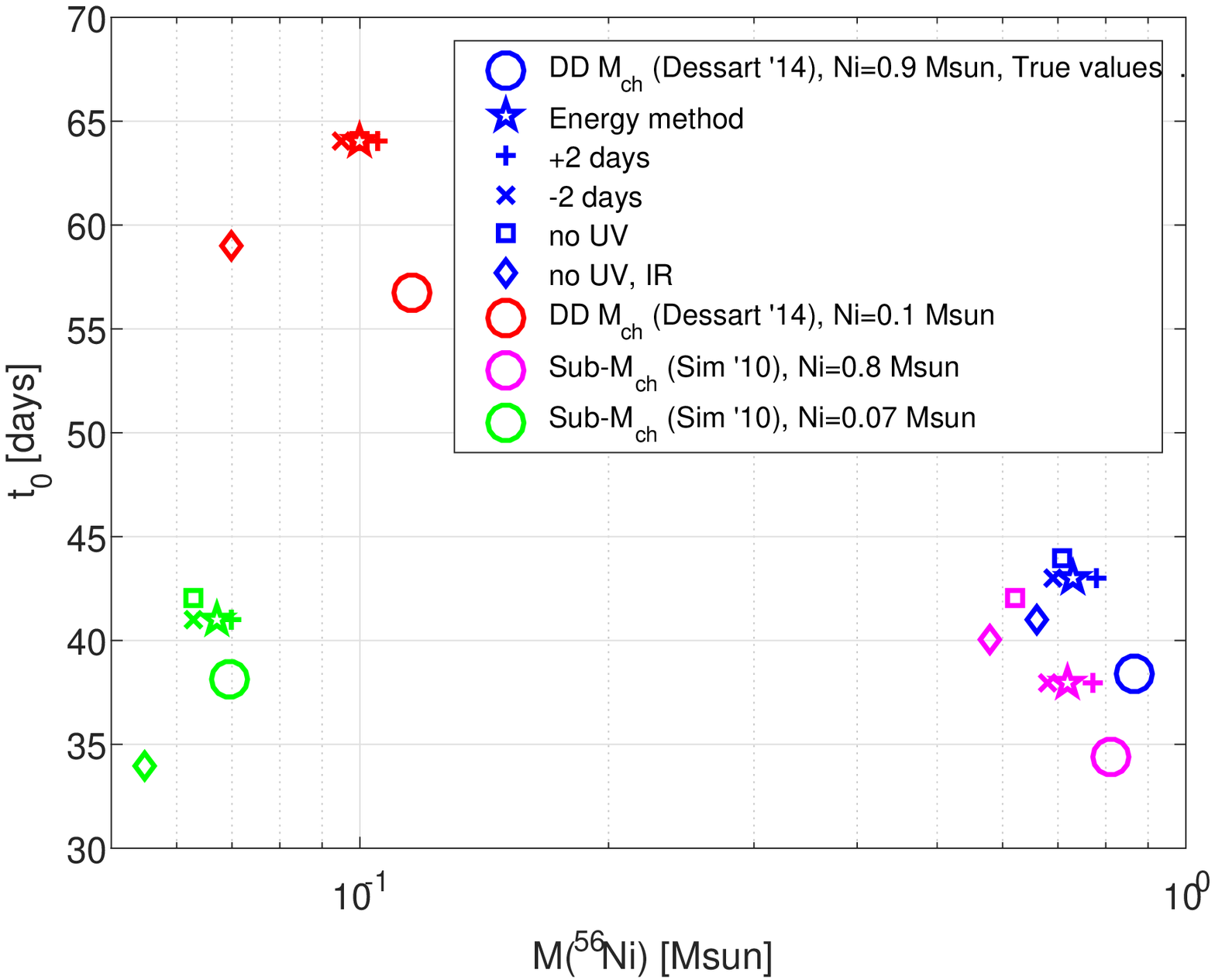}
	\caption{\textbf{Sensitivity of the $^{56}\rm Ni$ mass and $t_0$ to various effects.} The $t_0$ and $^{56}$Ni of the 4 ejecta presented in the top-left and bottom-left panels of figure \ref{fig:bolo_fit_t0}: high and low $^{56}\rm Ni$ mass for $M_{ch}$ models ejecta from \protect\cite{dbhk14} (blue and red, respectively) and high and low $^{56}\rm Ni$ mass for sub-$M_{ch}$ models ejecta from \protect\cite{s+10} (magenta and green, respectively). For each ejecta, the true value of $^{56}\rm Ni$ mass and $t_0$ is shown by a large circle, while the value inferred through Eqs. \eqref{eq:ratio} and (\ref{eq:qdep}-\ref{eq:qpos}), as shown in figure~\ref{fig:bolo_fit_t0}, is shown by a star. The values inferred by offsetting the lightcurves by $+2(-2)$ days are marked by $+(\rm x)$ signs, the values inferred by using bolometric lightcurves discarding the UV region ($\lambda<3000\angstrom$) are marked by squares and the values inferred by using bolometric lightcurves discarding the UV and also the IR region ($\lambda>10000\angstrom$) are marked by diamond signs.}\label{fig:bolo_fit_errors}
\end{figure}

\begin{figure*}
	\includegraphics[width=\columnwidth]{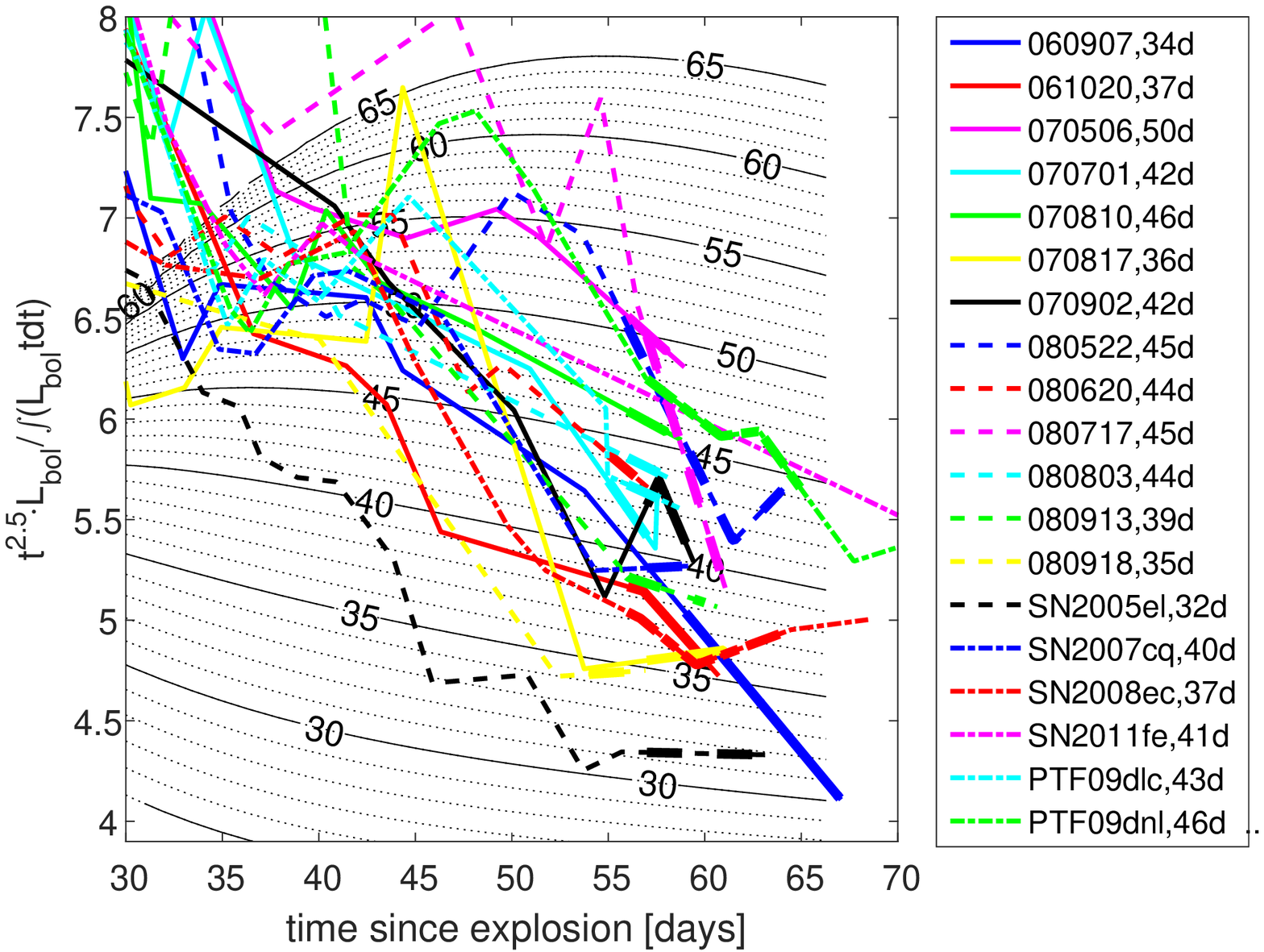}
	\hspace{2mm}
	\includegraphics[width=\columnwidth]{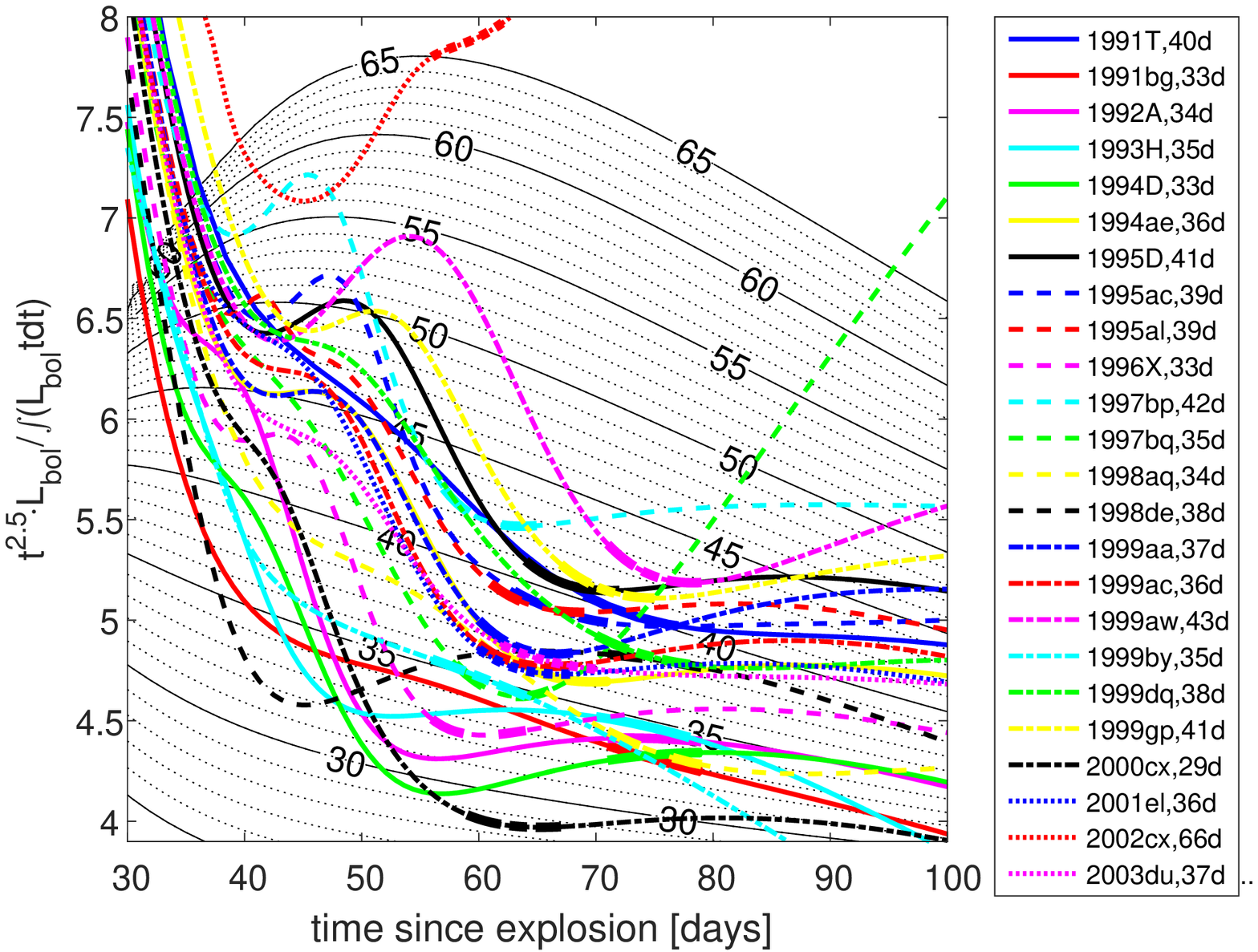}
	\caption{\textbf{left:} Fitting $t_0$ using bolometric lightcurves for the sample of \protect\cite{s+14}. The fitting function for each SN is compared to reference (black) lines with different $t_0$, as shown in equations (\ref{eq:ratio}, \ref{eq:qdep}-\ref{eq:qpos}). The legend shows, besides each SN, the fitted value of $t_0$, as inferred as the average of $t_0$ during the time range marked by the thicker region of each line. Note that for the first 13 plotted lightcurves, the first 5 characters were omitted from the legend as they're all identical: 'SNF20'. That is, for example, the full name of the first plotted lightcurve is SNF20060907. The inferred $^{56}\rm Ni$ masses, in order of their appearance in the legend, are as follows (in units of $M_{\odot}$): 
		           0.71, 0.36, 0.67, 0.78, 0.43, 0.36, 0.37, 0.67, 0.34, 0.79, 0.61, 0.48, 0.35, 0.49, 0.54, 0.37, 0.67, 0.54, 0.50.
		\textbf{right:} Fitting for $t_0$ as in the left panel but for the sample of \protect\cite{strit05}. The inferred $^{56}\rm Ni$ masses, in order of their appearance in the legend, are as follows (in units of $M_{\odot}$): 
		0.76, 0.066, 0.22, 0.19, 0.43, 0.68, 0.50, 0.77, 0.44, 0.57, 0.59, 0.53, 0.54, 0.055, 0.51, 0.54, 0.51, 0.07, 0.67, 0.58, 0.32, 0.31, 0.085, 0.29.}
	\label{fig:large_samples_fits}
\end{figure*}

We now turn to quantify the accuracy of inferring the ejecta parameters $t_0$ and $^{56}\rm Ni$ mass. The effect of various uncertainties on these parameters is shown in figure~\ref{fig:bolo_fit_errors} for the high and low $^{56}\rm Ni$ mass in Chandrasekhar and sub-Chandrasekhar mass explosions. The actual $^{56}\rm Ni$ mass and $t_0$ value (calculated using Eqs. \ref{eq:SigmaNi} and \ref{eq:t0exp}) are shown in the figure as large circles.

\textit{basic model accuracy}: 
The values that are obtained by comparing the synthetic light curves to Eqs. \eqref{eq:ratio} and (\ref{eq:qdep}-\ref{eq:qpos}) in the range $60{\rm d}<t<100{\rm d}$ are shown in the figure as large stars.    
As discussed above, Eqs. \eqref{eq:ratio} and (\ref{eq:qdep}-\ref{eq:qpos}) are accurate only up to several percent in the considered time range, causing the errors (overestimate) of $\sim10\%$ $t_0$ that are obtained. The accuracy in inferring the $^{56}\rm Ni$ mass given the correct value of $t_0$ is small ($\sim3\%$), but the systematic overestimation of $t_0$ causes a systematic underestimation of $^{56}\rm Ni$ mass by  $\sim10\%$ as can be seen in the figure.

\textit{explosion time}: The exact explosion time of observed supernovae is not known precisely.  The values of $t_0$ and $^{56}\rm Ni$ that are obtained by introducing errors of $\pm2$ days in the explosion time are shown as pluses and crosses. As can be seen, this has a negligible effect on the luminosity ratio and the deduced $t_0$. The $^{56}\rm Ni$ estimate is affected by $\pm5\%$ due to such errors in the explosion time.

\textit{Partial bolometric lightcurves}: The bolometric lightcurves of SNIa are reconstructed from observations in various regions of the spectrum. For this matter, the visible spectrum is usually taken into account, while UV ($\lambda < \sim3000\angstrom$) and IR ($\lambda > \sim10000\angstrom$) are not always available. While this is justified by the fact that most of the energy is emitted in the optical bands, it introduces some error. The values of $t_0$ and $^{56}\rm Ni$ mass that are deduced from the synthetic bolometric light curves when the UV (UV and IR) is not included are shown as squares (diamonds) in the figure and result in errors of up to $10\%$ in $t_0$. As can be seen, lack of IR can lead to a significant underestimate of the $^{56}$Ni mass by tens of percent.

\textit{Rising part of the lightcurve}: Finally, In many instances, the rising part of the lightcurve is only partially sampled. The error generated by this deficiency is conservatively estimated in the following way. The rising part of the light curves was replaced with one of two extreme possibilities: either  $L=L_{\rm peak}t^2/t_{\rm peak}^2$ (low) or  $L=L_{\rm peak}(1-(t-t_{\rm peak})^2/t_{\rm peak}^2)$ (high). These cases lead to changes of $\pm1$ day in the estimate of $t_0$, and of $\pm5\%$ in the deduced $^{56}\rm Ni$ mass for the 4 models considered.

\textit{Hydrodynamic effect of $^{56}\rm Ni$ decay}:
When $^{56}$Ni decays, it releases an energy per mass equal to about $(2500\rm km/s)^2$. In ejecta with $^{56}$Ni concentrated in the middle, the release is comparable to the velocity of the $^{56}$Ni and deviation from homologous expansion will occur \citep[e.g.][]{h+17}. To check how much this may affect our analysis, we ran a 1D hydrodynamic simulation for an ejecta with extreme $^{56}$Ni concentration: exponential velocity density profile with $M_{ch}$ mass and $2E/M=10^{18}\rm cm^2/s^2$ with the $^{56}$Ni concentrated in the center. We conservatively assumed that there is no diffusion in the radiation (but ignored the later $^{56}$Co decay that occurs when radiation diffuses significantly). The changes in the profiles caused the $\gamma$-ray escape time to change from $t_0=58$days (ignoring hydrodynamics) to $54$ days. The deviation from homologous expansion introduced an additional error of $\sim 2\%$ in eq.\eqref{eq:ratio}. Given these modest corrections in this extreme case, we conclude that the hydrodynamic effect of $^{56}$Ni decay can be ignored when inferring the bolometric parameters. 

Finally, we note that in addition to the above effects, observed lightcurves might suffer from additional problems such as jitter of measured fluxes and follow-up time shorter than needed. These effects are hard to quantify in general, as they depend on the details of each lightcurve. We further emphasize that  even though it was shown in several representing examples that we might systematically overestimate $t_0$ and underestimate the $^{56}\rm Ni$ mass, we do not correct for these possible biases in the following sections when fitting for observed light curves since we cannot be sure that this is a general feature for all ejecta.

\section{Type Ia's have a narrow range of gamma-ray escape times $t_0\approx 40\rm ~days$}\label{sec:observations}

In addition to the small sample shown on the right panel of figure~\ref{fig:bolo_fit_t0}, we fitted for the sample of 19 SNIa from the Nearby Supernova Factory analyzed in \cite{s+14} (which include near-IR corrections, but not UV corrections), as well as for the 24 lightcurves in the sample of \cite{strit05} that have low extinction (Galactic+host $E(B-V)<0.3$), i.e. the same sample as the one used in \cite{kk13}, which, as opposed to the Scalzo sample, contains several lightcurves with low $^{56}\rm Ni$ mass ($<0.3M_{\odot}$). The set of fitting curves for these samples with the reference lines to which they are compared is shown in figure~\ref{fig:large_samples_fits}, and the inferred value of $t_0$ for each SN is shown in the legend next to its name. For each SN, the $^{56}\rm Ni$ mass was inferred from the late time lightcurve using the inferred value of $t_0$. Table~\ref{table:sample} in appendix~\ref{sec:table} summarizes the determined values of $t_0$ and \nickel mass for all the bolometric light curves shown in this work. We note that the uncertainty of the $t_0$ values inferred from these light curves is larger than the values that were inferred for the light curves shown in the top right panel of figure~\ref{fig:bolo_fit_t0}. for the sample of \cite{strit05}, the inferred values of $t_0$ seem to reach a minimum and then rise significantly. This is probably due to the template that was used for the late times. But within the time range that was shown above, for well observed light curves, to be well suited for determining $t_0$, i.e. spanning 50-80 days, the uncertainty due to that is mostly limited to $\sim5$ days. For the sample of \cite{s+14}, several of the light curves have data only up to $\sim55$ days, which does not allow for much time when $t_0$ stabilizes on a constant value. In these cases, $t_0$ was determined from the last 3 days, but we warn the reader that these values are uncertain and might actually have been a bit lower had more data points been available.

\begin{figure}
	\includegraphics[width=\columnwidth]{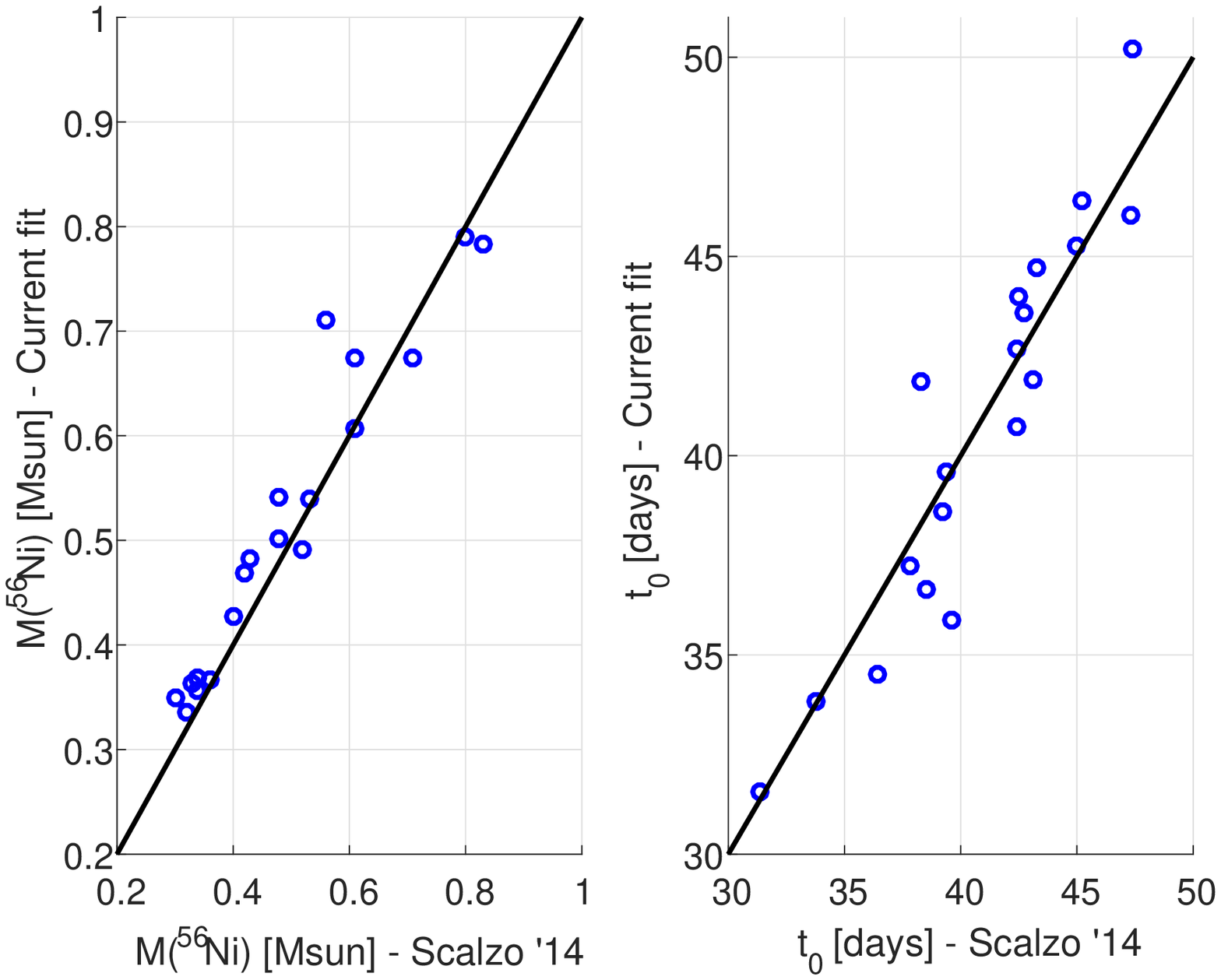}
	\caption{Comparing the inferred values of $^{56}\rm Ni$ mass (left panel) and of $t_0$ (right panel) as inferred by \protect\cite{s+14} and as obtained in this work.}
	\label{fig:t0_MNi_scalzo_us}
\end{figure}

A comparison for the values of $^{56}\rm Ni$ mass and $t_0$ obtained in this work and their values in \cite{s+14}, which were inferred using different methods, is shown in figure~\ref{fig:t0_MNi_scalzo_us}. Despite the different methods, we find similar results for the ejecta parameters: For $t_0$, the average value inferred here is 41 days, identical to the average of the inferred values in \cite{s+14}, and the average absolute difference is 1.3 days. 
For the $^{56}\rm Ni$ mass, the average of the sample in \cite{s+14} is $0.49M_{\odot}$, while the average value inferred here is $0.52M_{\odot}$ and the average difference between the methods is $0.04M_{\odot}$.

For the sample of \cite{strit05}, the values we infer for $^{56}\rm Ni$ mass are systematically $20\%-30\%$ lower than the estimate done by the author using Arnett's rule. This difference, when compared to the good agreement found for the \cite{s+14} sample, is due primarily to the different use of Arnett's rule by these two authors (the parameter, $\alpha$ was taken as $1$ by \cite{strit05} and as $1.2$ by \cite{s+14}), leading to a systematically higher estimate of $20\%$ by \cite{strit05}. Additionally, \cite{strit05} raised his fit for $^{56}\rm Ni$ mass by $10\%$ to compensate for the lack of UV and IR in the bolometric lightcurve, while we apply no such correction (as such a uniform correction was seen to not be consistent with simulations).

\section{Observations are consistent with sub-chandra and collision models across the range of luminosities and with Chandrasekhar models at the bright-end but not the faint end}\label{sec:models}

\begin{figure}
	\includegraphics[width=\columnwidth]{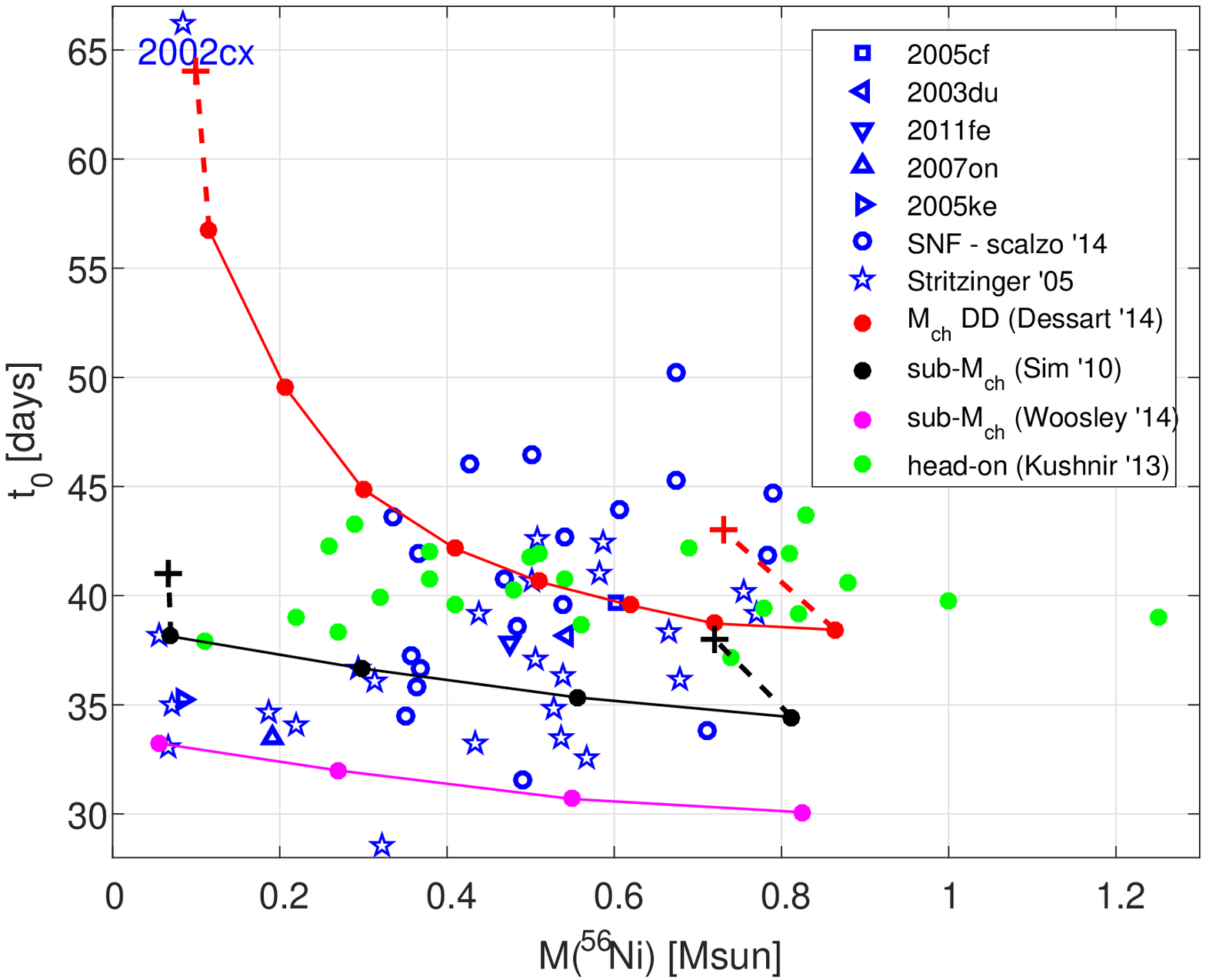}
	\caption{
		\textbf{Bolometric Width Luminosity Relation} The values of $t_0$ vs $ M(^{56}\rm Ni)$ ('width' as function of 'luminosity')  for a sample of SNe (blue marks, as detailed in the legend), and for different progenitor models: delayed detonation (red, \protect\cite{dbhk14}), two different sets of calculations of sub-Chandrasekhar central explosions (black - \protect\cite{s+10}, and magenta - S. Woosley presented in \protect\cite{mrkw14}) and head-on collisions (green - \protect\cite{kk13}). For 4 of the models, a dotted line is shown connecting the true ejecta values (marked by filled circles) to the values inferred by the fitting process (+ signs), indicating that a systematic small offset might be present between observed and true values.}
	\label{fig:MNi_vs_t0}
\end{figure}

As mentioned above, while the physics of SNIa lightcurves include complicated processes, one can constrain progenitor models independently of these processes using solely those ejecta parameters that can be inferred from the bolometric lightcurves. We compare the values found for $^{56}\rm Ni$ mass and $t_0$ in the previous section with those values in various progenitor models. 
The models we looked at were one dimensional delayed detonation models \citep{dbhk14}, one dimensional sub-Chandrasekhar central detonations of WDs by S. Woosley presented in \cite{mrkw14} (which do not result from explosion models, but serve as benchmarks for the simplest explosion configuration one can imagine, ejecta provided by S. Woosley), similar central detonations of sub-Chandrasekhar WDs from \cite{s+10}, and two dimensional head on collisions of CO WDs \citep{kk13}. The comparison of these models with the observed lightcurves presented in the previous section is shown in figure~\ref{fig:MNi_vs_t0}. 
The observations are consistent with the head-on collision model of \cite{kk13} (as already shown there) and with the sub-Chandrasekhar explosions of \cite{s+10} for the whole range of $^{56}\rm Ni$ masses. They are in possible tension with the sub-Chandrasekhar explosions of Woosley at the high end of the $^{56}\rm Ni$ masses, though due to the large observational scatter they're not inconsistent with it, and in addition, the fact that we showed that our fitting method likely overestimates $t_0$ could resolve some of this tension.
The observations are inconsistent with delayed detonations of Chandrasekhar mass WDs as an explosion mechanism for the full range of observed luminosities, since these give values of $t_0>50$ days for $^{56}\rm Ni$ masses below $0.2 M_{\odot}$, whereas the observational values of $t_0$ are all lower than this, even more in the range of such low $^{56}\rm Ni$ masses (the one observational exception, sn2002cx, is known to be an extremely peculiar, non representative case of SNIa \citet{li+03})
Moreover, if we take into account the likely overestimation of $t_0$ in our fitting procedure, the inconsistency becomes even more significant. The observations are consistent with Chandrasekhar mass models at the bright end of the luminosity range.

\begin{figure}
	\includegraphics[width=\columnwidth]{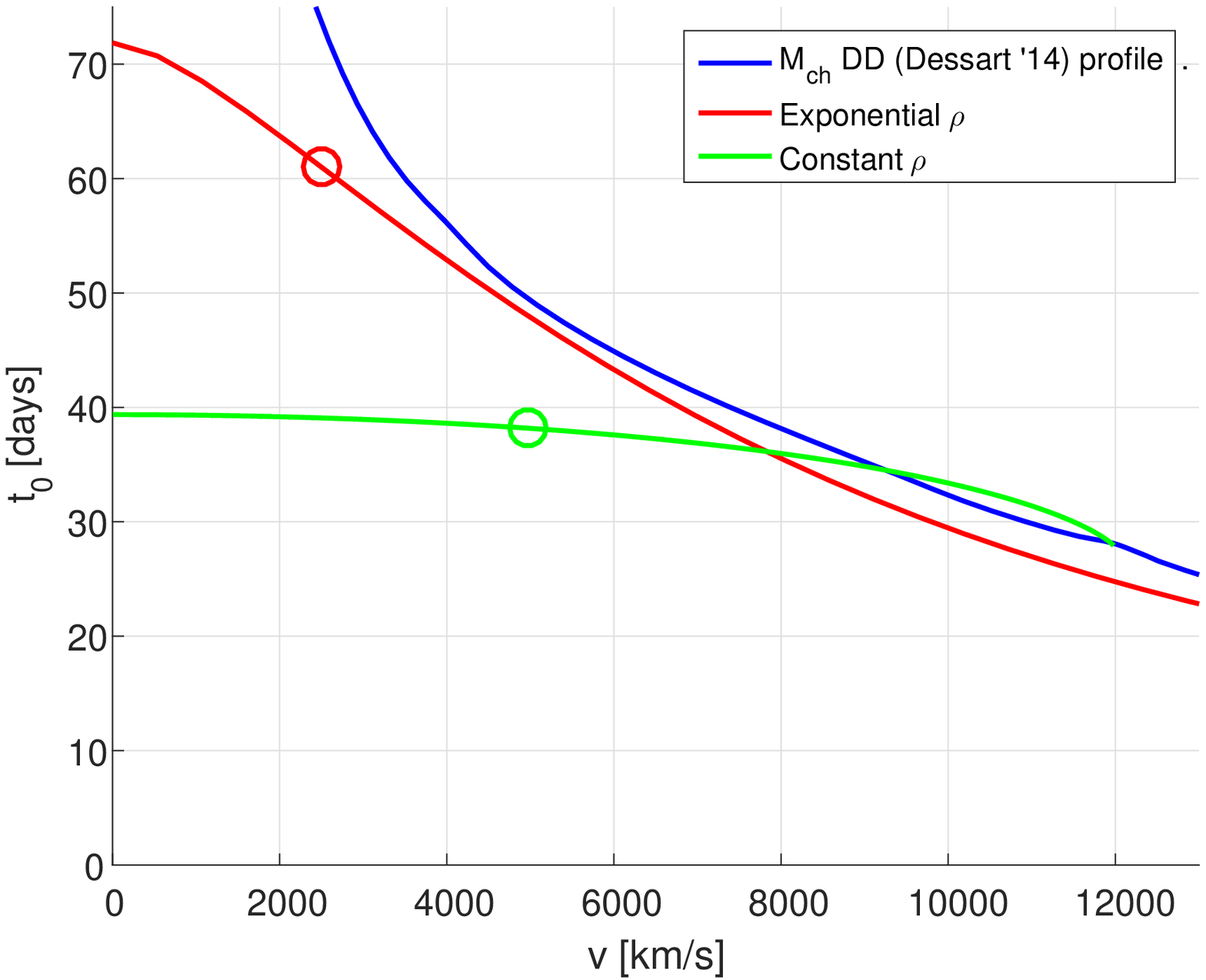}
	\caption{$t_0$ as a function of the location of the $^{56}\rm Ni$ within the ejecta for various density profiles (assuming it is concentrated in a thin shell): the lowest $^{56}\rm Ni$ mass delayed detonation from \protect\cite{dbhk14} (blue line), exponential and constant profiles (red and green). The circles mark the velocity at the edge of the inner $0.1M_{\odot}$.}
	\label{fig:t0_vs_vel}
\end{figure}

\begin{figure}
	\includegraphics[width=\columnwidth]{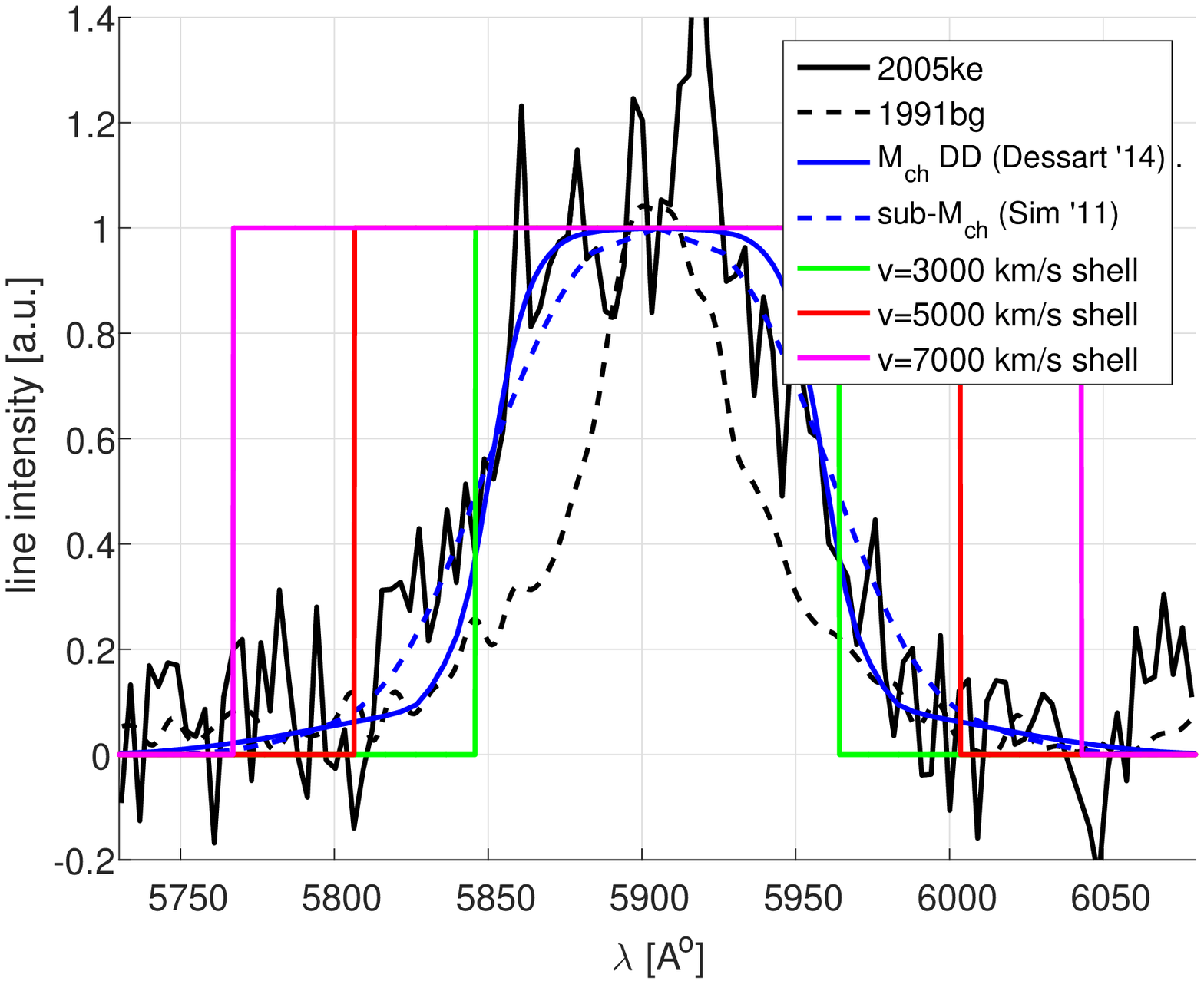}
	\caption{Emission line of $^{56}\rm Co$ at $\sim5900\angstrom$ observed in two low $^{56}\rm Ni$ mass SN: sn1991bg \protect\citep{t+96} and sn2005ke \protect\citep{f+13} (dashed and solid black lines), along with line-of-sight $^{56}$Ni mass distribution, serving as indicators for line shapes, for low $^{56}\rm Ni$ mass ejecta for delayed detonations \protect\citep{dbhk14} and central sub-Chandrasekhar WD explosions \protect\citep{s+10} (solid and dashed blue lines), as well as line shapes for thin shells of $^{56}\rm Co$ at velocities of 3000, 5000 and 7000 km/s (green, red and magenta solid lines, respectively).}
	\label{fig:lines_profile}
\end{figure} 

\section{Summary and discussion}\label{sec:discussion}
In this paper we derived and applied a method to derive the gamma-ray escape time $t_0$ and the $^{56}$Ni mass 
from bolometric light-curves using energy conservation from basic principles. The method is derived in  \S\ref{sec:physics}, validated using detailed radiation transfer calculations in \S\ref{sec:simulations} and applied to observations in \S\ref{sec:observations} (see in particular figures \ref{fig:bolo_fit_t0} and \ref{fig:MNi_vs_t0}). We find that type Ia's at all luminosities have $t_0$ values in a narrow range around $30-45$ days with weak or no correlation with $^{56}Ni$ mass.

When applied to the Supernova Factory sample in \citep{s+14} we obtain similar results for $M_{\rm Ni56}$ and $t_0$ as those of \citep[][see figure \ref{fig:t0_MNi_scalzo_us}]{s+14}. \citet{s+14} used a  sophisticated code that involved comparison to detailed models and is based on Arnett's rule. We emphasize that the method presented here is different, simpler and requires less assumptions. Perhaps a more important difference between the two works is that \citet{s+14} did not compare the resulting values of $t_0$ to those of models directly as presented here in figures \ref{fig:bolo_fit_t0} and \ref{fig:MNi_vs_t0}. Instead they tried to deduce the ejecta mass from the values of $t_0$ and $^{56}$Ni by using explosion simulations for calibration and compared the deduced masses to those of various models. This last step depends on the calibration model and may lead to misleading results. In fact, different models with different ejecta masses can have very similar $t_0$ and $^{56}$Ni. For example, the collision models and the sub-$M_{ch}$ have similar values of $t_0$ but significantly different ejecta masses at the high $^{56}$Ni end. In particular the collision of two $0.9$ $M_{\odot}$ WDs (total mass of $1.8 M_{\odot}$, $t_0=39$d, $^{56}$Ni$=0.79 M_{\odot}$) and the central detonation of a WD with mass of $M=1.15 M_{\odot}$ ($t_0=35$d, $^{56}$Ni$=0.81 M_{\odot}$) have very similar $t_0$ and $^{56}$Ni while having masses that differ by more than 50 \%. \textbf{The total mass cannot be extracted from $t_0$ and the $^{56}$Ni mass. However, it can easily be directly compared to the values of any multi-D explosion model}. Finally we note that the sample in \cite{s+14} does not include low luminosity type Ia's. As can be seen in figure \ref{fig:MNi_vs_t0}, the different models deviate most significantly at the faint end and thus low luminosity type Ia's, although constituting only a small fraction of the total observed SNIa, play a significant role in differentiating between different explosion mechanisms. that may explain the full range of luminosities. We note that it is also possible that different mechanisms produce type Ia supernovae with different luminosities, and Chandrasekhar mass models are not disfavored based on their bolometric properties as progenitors for bright type Ia’s. 

The $^{56}$Ni and $t_0$ deduced from the observations are compared to models in figures \ref{fig:bolo_fit_t0} and \ref{fig:MNi_vs_t0}. We find that low luminosity supernovae are inconsistent with delayed-detonation Chandrasekhar mass models in agreement with the findings of several recent works \citep{kk13,s+14,s+14b,b+17,dhawan17}.

The large escape times for low $^{56}\rm Ni$ masses in $M_{ch}$ models, are due to the large mass present at velocities beyond the $^{56}\rm Ni$.
Given the uncertainty of the Chandrasekhar mass explosion models, it is instructive to ask if the problem can be elevated by changing the distribution of the $^{56}$Ni in the ejecta. As can be seen in figure \ref{fig:t0_vs_vel}, the $^{56}$Ni has to be moved out to velocities of $\sim7000$ km/s in order to reduce the escape time to observed values of $t_0\sim 40$ days. Such high $^{56}$ Ni velocities are inconsistent with nebular spectra of low luminosity type Ia supernovae \citep[e.g.][and figure \ref{fig:lines_profile}]{m+98} which imply velocities $v\lesssim 5000$km/s. We note that in an extreme case of a Chandrasekhar-mass ejecta with a homogeneous density distribution, both the line-width and the bolometric properties may be accounted for. Such ejecta are not expected in thermonuclear explosions as far as we know.  Recently \cite{h+17} showed results of a series of radiation transfer calculations that are claimed to agree with type Ia observations across the type Ia brightness range. Based on the results here, either the bolometric luminosity or the nebular line widths of these models (both are not shown in the paper) are likely inconsistent with observations of low-luminosity type Ia's.

Direct collisions and sub-Chandrasekhar models have a narrow range of $t_0$ in agreement with observations. The existence of such examples, and in particular the collisions where the ignition of a detonation is robust and resolved in global simulations, supports the existence of a single mechanism for type Ia supernovae that explains the entire range of brightness. 

\section*{acknowledgments}
We thank Doron Kushnir, Subo Dong, Chris Burns, Mark Phillips, Stuart Sim, Stephane Blondin, Stan Woosley, Bob Kurucz, and Peter Hoeflich for useful discussions and clarifications. We thank Charling Tao and Richard Scalzo for help with the bolometric light curves of the Supernova Factory. We thank Stuart Sim, Stephane Blondin, Stan Woosley and Daniel Kasen for providing ejecta and radiation transfer results. This research was supported by the ICORE Program (1829/12) and the Beracha Foundation.

\appendix

\section{Attached code and data files}\label{sec:files}
There are two .zip files attached to this paper in the online supplementary material which can also be accessed using the following dropbox links:\\ 
AnalyzeLightCurves.zip at \\
https://www.dropbox.com/s/ngsnhwof3ron5og/ \\AnalyzeLightCurves.zip?dl=0\\
AnalyzeModels.zip at \\
https://www.dropbox.com/s/5mcu1jy7p5n0kyw/ \\AnalyzeModels.zip?dl=0\\ 
Each of these compressed files contains a folder with additional files and folders as described below. 
\subsection{Light Curve Analysis Files}
The file  AnalyzeLightCurves.zip, contains a folder with matlab and python Jupyter notebooks that allow the analysis of bolometric light curves along with data files of the supernovae presented in figure \ref{fig:bolo_fit_t0} as important examples. The following files are included in the folder:
\begin{enumerate}
	\item Bolometric UVOIR light curves of the 5 well observed SNe presented in the top right panel of figure \ref{fig:bolo_fit_t0}. The references are described in the caption and the explosion time estimates are explained in the header of the files. The five files are: sn2003du\_Stanishev07.txt, sn2005cf\_Wang09.txt, sn2005ke\_Phillips12.txt, sn2007on\_Phillips12.txt, sn2011fe\_Mazzali15.txt. Each file contains two columns- time in days and Log10 of the luminosity in ergs per second.
	\item SNIa\_t0\_from\_bolometric.m : A matlab file that creates the top right panel of figure \ref{fig:bolo_fit_t0} and can be used to extract the value of $t_0$ for a given bolometric light curve. It uses the attached light curve files and thus needs to be run from the folder in which they are stored.
	\item SNIa\_t0\_MNi\_from\_bolometric.ipynb : A python Jupyter notebook that allows the extraction of $t_0$ and the $^{56}$ Ni  mass from bolometric light curves by manual fitting based on equations (\ref{eq:ratio}, \ref{eq:qdep}-\ref{eq:qpos}). This file also uses the (same) bolometric data files and needs to be run from the folder where they are stored. 
\end{enumerate} 

\subsection{Model Analysis Files}
The file AnalyzeModels.zip contains a folder with files for analyzing model ejecta and performing Monte Carlo gamma-ray transfer calculations. It contains the following files and folders.
\begin{enumerate}
	\item There are three folders with ejecta files containing the total density and the $^{56}$Ni distribution for the models described in \S\ref{sec:repmodels}\\
	** DDC\_Blondin13/ : $M_{ch}$ 1D delayed detonation models from \cite{dbhk14}\\
	** SubChandraWoosley14/ : 1D sub-$M_{ch}$ central detonations of WDs with different masses by S. Woosley and presented in \citep{mrkw14}\\
	** Col2DKushnir13/: 2D direct collision models  from \cite{kk13}. Since these 2D ejecta files take up some memory, only equal mass collisions are provided. All ejecta from  \cite{kk13} in a similar format can be accessed in the following dropbox link:\\
	https://www.dropbox.com/sh/7g25m6zcv4nva2u/\\AABQwpExS\_5JZNE68-lsRBbTa?dl=0\\
	The files of these models are read by the python Jupyter notebook described next.
	\item SNIa\_Analyze\_Models.ipynb : A python Jupyter notebook that allows the extraction of $t_0$ and the calculation of gamma-ray transfer for model ejecta. In particular this file reads and allows the analysis of the ejecta in the model folders above and needs to be run from a folder that contains the model folders. The file allows gamma-ray transfer calculations using a c-based code which is attached in the folder MonteCarloCode. In order for the Jupyter notebook to access the Monte Carlo code, the MonteCarloCode\ folder needs to be stored at the same folder as the notebook.
	\item MonteCarloCode\ : A folder that contains a gamma-ray transfer Monte Carlo code written in c. It can be run directly but is easiest to use with the Jupyter notebook. The main outputs are log\_output.txt which contain $t_0$ and output.txt that contains the total deposition by gamma-rays and positrons as a function of time. The provided examples are 1D and 2D, but the code contains functions for calculating the 3D ejecta. 
\end{enumerate}

\section{$t_0$ and \nickel mass from light curves}\label{sec:table}
Table~\ref{table:sample} summarizes the sample of bolometric light curves analyzed in this work, along with the inferred values of $t_0$ and \nickel mass. Note that SN2011fe appears both in \cite{m+15} and in the sample of \cite{s+14}, yielding similar results, and SN2003du appears both in \cite{s+07} and in \cite{strit05}, yielding similar results for $t_0$ but different results for the \nickel mass, due to vastly different distances to the source used in these two.

\begin{table}
	\caption{observed SN sample.}
	\label{table:sample}
	\begin{tabular}{|c|c|c|c|}
		\hline 
		name & $t_0$ [days] & \nickel [$M_{\odot}$] & ref for Bolometric \\ 
		\hline 
		2005cf & 40 & 0.60 &  \cite{w+09} \\ 
		\hline 
		2003du & 38 & 0.54 &  \cite{s+07}  \\ 
		\hline 
		2011fe & 38 & 0.47 &  \cite{m+15}  \\ 
		\hline 
		2007on & 34 & 0.19 & \cite{phil12}  \\ 
		\hline 
		2005ke & 35 & 0.085 & \cite{phil12}  \\ 
		\hline 		
		1991T & 40 & 0.76 &  \cite{strit05}  \\ 
		\hline 
		1991bg & 33 & 0.066 &  \cite{strit05}  \\ 
		\hline 
		1992A & 34 & 0.22 &  \cite{strit05}  \\ 
		\hline 
		1993H & 35 & 0.19 & \cite{strit05} \\ 		
		\hline 
		1994D & 33 & 0.43 & \cite{strit05} \\ 
		\hline 
		1994ae & 36 & 0.68 & \cite{strit05}  \\ 
		\hline 
		1995D & 41 & 0.50 & \cite{strit05} \\ 
		\hline 
		1995ac & 39 & 0.77 &  \cite{strit05}  \\ 
		\hline 
		1995al & 39 & 0.44 &  \cite{strit05}  \\ 
		\hline 
		1996X & 33 & 0.57 & \cite{strit05}  \\ 		
		\hline 
		1997bp & 42 & 0.59 & \cite{strit05} \\ 		
		\hline 
		1997bq & 35 & 0.53 & \cite{strit05} \\ 		
		\hline 						
		1998aq & 34 & 0.54 & \cite{strit05} \\ 
		\hline 
		1998de & 38 & 0.055 &  \cite{strit05} \\ 
		\hline
		1999aa & 37 & 0.51 & \cite{strit05}  \\ 		
		\hline 
		1999ac & 36 & 0.54 & \cite{strit05} \\ 		
		\hline
		1999aw & 43 & 0.51 & \cite{strit05} \\ 		
		\hline 		 				 
		1999by & 35 & 0.070 & \cite{strit05} \\ 
		\hline
		1999dq & 38 & 0.67 & \cite{strit05} \\ 		
		\hline
		1999gp & 41 & 0.58 & \cite{strit05}  \\ 		
		\hline 		 		 
		2000cx & 29 & 0.32 &  \cite{strit05} \\ 
		\hline 
		2001el & 36 & 0.31 &  \cite{strit05} \\ 
		\hline
		2002cx & 66 & 0.085 & \cite{strit05} \\ 		
		\hline 
		2003du & 37 & 0.29 & \cite{strit05} \\ 		
		\hline 	
		SNF20060907-000 & 34 & 0.71 & \cite{s+14} \\ 		
		\hline 						 
		SNF20061020-000 & 37 & 0.36 & \cite{s+14} \\ 		
		\hline 						 
		SNF20070506-006 & 50 & 0.67 & \cite{s+14} \\ 		
		\hline 						 
		SNF20070701-005 & 42 & 0.78 & \cite{s+14} \\ 		
		\hline 						 
		SNF20070810-004 & 46 & 0.43 & \cite{s+14} \\ 		
		\hline 						 
		SNF20070817-003 & 36 & 0.36 & \cite{s+14} \\ 		
		\hline 						 
		SNF20070902-018 & 42 & 0.37 & \cite{s+14} \\ 		
		\hline 						 
		SNF20080522-011 & 45 & 0.67 & \cite{s+14} \\ 		
		\hline 						 
		SNF20080620-000 & 44 & 0.34 & \cite{s+14} \\ 		
		\hline 						 
		SNF20080717-000 & 45 & 0.79 & \cite{s+14} \\ 		
		\hline 						 
		SNF20080803-000 & 44 & 0.61 & \cite{s+14} \\ 		
		\hline 						 
		SNF20080913-031 & 39 & 0.48 & \cite{s+14} \\ 		
		\hline 						 
		SNF20080918-004 & 35 & 0.35 & \cite{s+14} \\ 		
		\hline 						 
		SN2005el & 32 & 0.49 & \cite{s+14} \\ 		
		\hline 						 
		SN2007cq & 40 & 0.54 & \cite{s+14} \\ 		
		\hline 						 
		SN2008ec & 37 & 0.37 & \cite{s+14} \\ 		
		\hline 						 
		SN2011fe & 41 & 0.47 & \cite{s+14} \\ 		
		\hline 						 
		PTF09dlc & 43 & 0.54 & \cite{s+14} \\ 		
		\hline 						 
		PTF09dnl & 46 & 0.50 & \cite{s+14} \\ 		
		\hline 						 
	\end{tabular} 
\end{table}


\bibliographystyle{mnras}
\bibliography{jets_bib}

\bsp
\label{lastpage}
\end{document}